# Lessons Learned from Teaching Astronomy with Virtual Reality

Philip Blanco[1,2], Gur Windmiller[2], William Welsh[2], and Sean Hauze[2]

[1]*Grossmont College, 8800 Grossmont College Drive, El Cajon, CA 92020*

[2]*San Diego State University, 5500 Campanile Drive, San Diego, CA 92182*

**Abstract** We report on the initial phase of an ongoing, multi-stage investigation of how to incorporate Virtual Reality (VR) technology in teaching introductory astronomy concepts Our goal was to compare the efficacy of VR vs. conventional teaching methods using one specific topic - Moon phases and eclipses. After teaching this topic to an ASTRO-101 lecture class, students were placed into 3 groups to experience one of three additional activities: supplemental lecture, "hands-on" activity, or VR experience. All students were tested before and after their learning activity. Although preliminary, our results can serve as a useful guide to expanding the role of VR in the astronomy classroom

## 1. Virtual Reality for Astronomy Education

Many topics in astronomy are difficult for students to grasp because of their vast scales and complex relationships in space and time. The recent proliferation and reduced cost of immersive technologies such as Virtual Reality (VR) may provide astronomy educators with new opportunities for effectively conveying these concepts.

VR technology not only allows the display of astronomical scenarios in 3-D, but also immerses the student in them, allowing each student to take control and explore ideas by interacting with objects and changing their viewpoints.

### 1.1. Potential Advantages

An appealing aspect of VR is the possibility of unstructured "play" learning. Students can move around the virtual environment and take part in unscripted constructivist discovery, with immediate feedback for "What if...?" questions. This models the scientific method in addition to providing memorable experiences, which can be drawn upon when the student is tested.

### 1.2. Potential Concerns

Our major concern was the time cost involved, for our students and ourselves. We knew we would have to develop our own training materials for use of the VR hardware, and for the astronomical topic. Students would have to first learn how to use the system, and then work through the VR "scenario" that we had developed. In contrast to teaching a lecture to students "in parallel", we were limited by staff and hardware to working with a few students at a time, "in series". A related concern was that the technology itself would become a distraction for students, and inhibit their learning of the material.

A final concern to be addressed is how to effectively assess whether tasks were completed, and learning occurred, in the VR environment. For example, is it fair to give a multiple-choice test in "flat-land", when a student has demonstrated their spatial and temporal understanding in a 3-d immersive environment?

## 2.   Our Investigation

To explore the efficacy of immersive technology in the classroom, in 2018 Spring we tested VR vs. traditional teaching methods in an ASTRO 101 class at San Diego State University. Since no "package" of VR astronomy lessons existed, we developed our own interactive VR activity for one topic: Moon phases and eclipses. This topic lends itself well to 3-dimensional (3D) visualization and interaction, and yet can be problematic for students to learn (Chastenay 2016, and references therein).

Since we did not know whether the VR activity would put students at a disadvantage when it came to exams, we taught all students an initial lecture on Moon phases and eclipses; students also had access to the textbook's treatment of this topic. After this standard lecture, the class was divided into three cohorts (one "VR" and two "control") for further activities with the purpose of strengthening their understanding of the topic.

### 2.1.   The VR Activity

We used the *Universe Sandbox 2* software package to develop our own VR activity to teach Moon phases and eclipses. This software allowed us to create some precursor tasks for the student to get used to the VR goggles and controllers, and then provide them with an interactive scenario that modeled the Sun, Earth and Moon. The student would interact with these objects to perform tasks and answer a series of questions.

In an attempt to reduce time overhead, and to provide some guidance, we adopted a "buddy system" strategy from another "immersive" technology - SCUBA diving - with students working in pairs. One student puts on the VR head-mounted display (HMD), while the other reads instructions and observes on a flat screen what the VR student is seeing. Then each student switches places. The total time using the HMD was limited to 20 minutes.

### 2.2.   Control Activities

Two cohorts of students provided control groups for our investigation. One group received a 20-minute additional lecture with new diagrams and animations. The other control group participated in a "hands-on" activity, using a light source in the classroom to model sunlight, and foam balls to represent the Moon. This activity was based on that provided in *The Universe at your Fingertips* guide developed for Project ASTRO (Fraknoi 2011) The activity was also limited to 20 minutes, so that students in all three cohorts were given the same amount of instruction time.

### 2.3. Assessment

All students were tested pre- and post-activity using multiple-choice questions on Moon phases and eclipses, drawn primarily from the *Astronomy Diagnostic Test* (Zeilik 2002) *Private Universe*, and University of Nebraska, Lincoln *ClassAction* question banks (Lee 2010). The questions were chosen to be in the "difficult" category, requiring students to apply their spatial and temporal reasoning. We also asked students to provide feedback on the VR activity using both Likert scale and freeform survey questions, adapted from John Keller's Instructional Materials Motivation Survey (IMMS) measuring student motivation to learn through the lens of the attention, relevance, confidence, and satisfaction (ARCS) model (Keller 1987).

### 3. Tentative Results

We consider our results as tentative because, in hindsight, we realized that the composition of our student cohorts was not uniform. The largest source of bias resulted from allowing students to volunteer for the VR activity. We also did not track separately those students who were taking a separate astronomy laboratory course, which possibly (but not necessarily) included a laboratory activity devoted to lunar cycles.

The pre- and post-activity scores of the three cohorts are shown in Figure 1. With the caveats above, we conclude that VR seemed to be at least as effective as the other additional activities. In other words, VR did not "hurt" the students' understanding of Moon phases and eclipses, as tested in a multiple-choice exam.

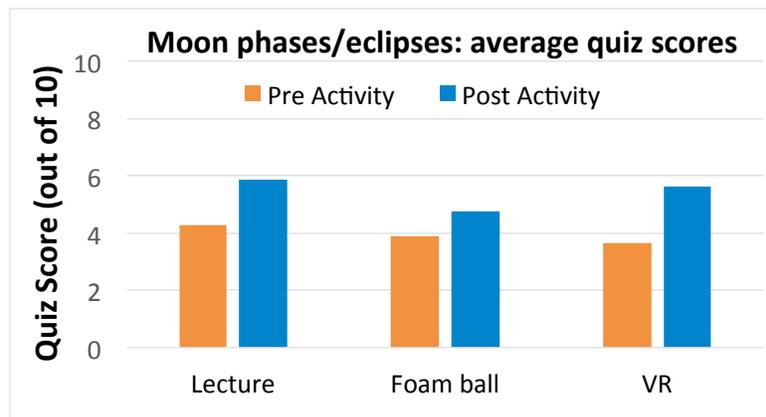

Figure 1. Pre- and post- activity quiz scores for the three supplemental activities.

We were also interested in which students found the VR activity most helpful, so despite the small sample sizes, we split each cohort into two groups by course grade. From this we tentatively conclude that the students who benefitted the most were in the VR cohort and in the upper 50% of the class.

From the survey questions, we found that most students reacted positively to the VR experience and felt that it contributed to their learning. However, even "tech-savvy" students appreciated the need for clear and concise instructions on the use of the VR system.

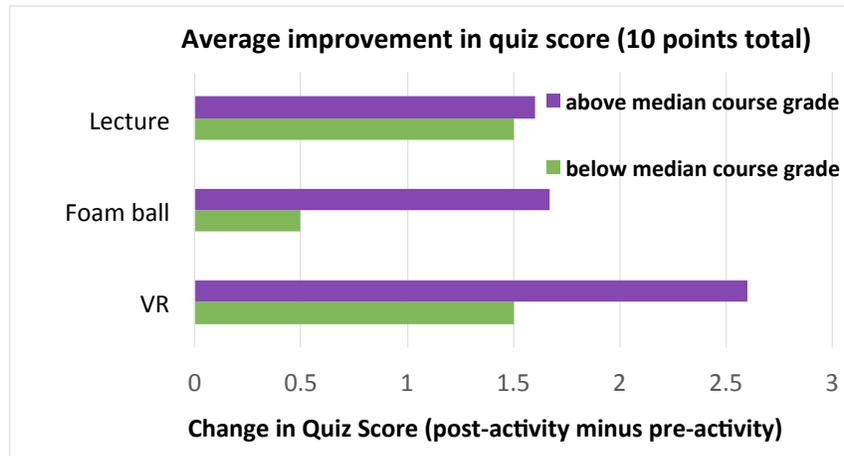

Figure 2. Comparison of score improvements for those students above (purple) or below (green) the median overall course grade.

### 4.   Next steps

From this pilot study, we are now confident that we can use VR as the primary instruction method for this topic, and in the near future we shall test its efficacy against control groups without the prior "traditional" lecture. The overhead of training can be reduced in a lecture or laboratory class by offering more VR activities, which we would like to develop using *Universe Sandbox 2* or other software. Possible topics amenable to VR include parallax, binary stars, Hubble expansion, and the Doppler effect. A major limitation is the time needed to develop these VR scenarios, even before writing a "scripted" lesson plan. We anticipate that these tasks will become easier as user-friendly software for creating virtual scenarios is developed.

### References

Chastenay, P. 2016, "From geocentrism to allocentrism: teaching phases of the Moon in a digital full-dome planetarium," *Research in Science Education.*, 46, 43

Keller, J. M. 1987, "Development and use of the ARCS model of instructional design," *Journal of Instructional Development,* 10, 2

Lee, K. 2010, "A multi-institution study on the effectiveness of ClassAction to promote student understanding in Astro 101," *Bulletin of the American Astronomical Society,* 42, 415

The Universe at Your Fingertips 2.0 DVD-ROM. A. Fraknoi (ed). Astronomical Society of the Pacific (San Francisco, CA, 2011).

Zeilik, M. 2002, "Birth of the Astronomy Diagnostic Test: prototest evolution," *Astronomy Education Review*, 1, 46